\documentclass[namedreferences]{solarphysics}
\usepackage[hyperref,optionalrh,solaromanenum]{spr-sola-addons} 
\usepackage{graphicx}                    
\usepackage{color}                       

\usepackage{amsmath, bm}

\begin{document}

\begin{article}

\begin{opening}

\title{The Sun's Non-Potential Corona over Solar Cycle 24}

%
\author[addressref={aff1},corref,email={anthony.yeates@durham.ac.uk}]{\inits{A.R.}\fnm{Anthony R. }\lnm{Yeates}\orcid{}}

%

\address[id=aff1]{Department of Mathematical Sciences, Durham University, Durham, DH1 3LE, UK}

\begin{abstract}
The global magnetic field in the solar corona is known to contain free magnetic energy and magnetic helicity above that of a current-free (potential) state. But the strength of this non-potentiality and its evolution over the solar cycle remain uncertain. Here we model the corona over Solar Cycle 24 using a simplified magneto-frictional model that retains the magnetohydrodynamic induction equation but assumes relaxation towards force-free equilibrium, driven by solar surface motions and flux emergence. The model is relatively conservative compared to some others in the literature, with free energy approximately $20-25\%$ of the potential field energy. We find that unsigned helicity is about a factor 10 higher at Maximum than Minimum, while free magnetic energy shows an even greater increase. The cycle averages of these two quantities are linearly correlated, extending a result found previously for active regions. Also, we propose a practical measure of eruptivity for these simulations, and show that this increases concurrently with the sunspot number, in accordance with observed coronal mass ejection rates. Whilst shearing by surface motions generates $50\%$ or more of the free energy and helicity in the corona, we show that active regions must emerge with their own internal helicity otherwise the eruptivity is substantially reduced and follows the wrong pattern over time.
\end{abstract}

%

\end{opening}

\section{Introduction}\label{s:intro} 

There is little doubt that the Sun's coronal magnetic field does not simply evolve through a sequence of current-free equilibria as assumed by the traditional Potential Field Source Surface, or PFSS model \citep{altschuler1969, schatten1969}. Indirect evidence for low coronal currents includes the presence of sheared filament channels \citep[e.g.,][]{martin1998,mackay2010, sheeley2013}, the twisted shape of extreme-ultraviolet or X-ray loops \citep[e.g.,][]{pevtsov2002, malanushenko2014}, and the presence of coronal flux rope cavities \citep[e.g.,][]{ruminska2022}.

On theoretical grounds one expects free magnetic energy -- above that of a current-free configuration -- to build up only in regions of closed magnetic field. This is because when an open field region is sheared by moving the field line footpoints on the solar surface, the lack of line-tying at the outer end means that this open field is free to relax back to a minimum-energy (current-free) configuration. But in a closed-field region the field lines are line-tied on the photosphere at both ends and will retain the shear.

What prevents free energy building up indefinitely in the corona? There are three factors to consider \citep{mikic1994}:
\begin{enumerate}
    \item As coronal arcades are sheared, they expand and more of the field becomes open, subsequently relaxing to potential and thus releasing energy.
    \item If strong enough current sheets form in the corona then these can support magnetic reconnection allowing reconfiguration of the magnetic field to a lower energy state (and possibly also the loss of magnetic energy through ohmic heating). This can be gradual or sudden.
    \item If the magnetic field becomes too sheared or twisted, it can lose equilibrium leading to an eruption that removes energy through the outer boundary. This is thought to be the origin of coronal mass ejections.
\end{enumerate}

One can obtain a feeling for how the free energy varies over the solar cycle by looking at the observed rates of solar flares \citep[e.g.,][]{hathaway2015} or coronal mass ejections \citep{webb2012, lamy2019}. These indices of solar activity tend to follow (roughly) the observed sunspot number, because most of these events originate from active regions. For active regions, non-potential magnetic field extrapolations are now routinely made \citep{wiegelmann2021}, giving local estimates of free energy albeit with some uncertainty \citep[e.g.,][]{regnier2007,derosa2015,thalmann2022}. But the magnetic flux from active regions spreads out gradually as they decay and interact, and the effect of this process on free energy is harder to pin down. Indirect observations show that there is indeed free energy stored outside of active regions, in concentrated filament channels or even large-scale sheared arcades. But to date there have been relatively few quantitative studies of how the free energy in the global solar corona varies over the solar cycle, two exceptions being \citet{yeates2010} and \citet{chifu2022}.

\citet{chifu2022} used static nonlinear force-free field (NLFFF) extrapolations from synoptic vector magnetograms, once per 27-day Carrington rotation, to follow the evolution of the coronal magnetic field over Solar Cycle 24 (from 2010 to 2019). They found a peak of free energy of about $2.5\times 10^{33}\,\mathrm{erg}$ -- occurring in late 2014 which is the time when observed magnetic flux and flare activity peaked for that cycle. These authors did not indicate the relative magnitude of this free energy compared to the current-free field; however, for a single similar extrapolation, \citet{tadesse2014} previously obtained a figure of $15\%$. On the other hand, this earlier work illustrated -- as confirmed in the model comparison study by \citet{yeates2018} -- that these global NLFFF extrapolations from present-day vector magnetogram observations are unable to accurately recover currents outside active regions, such as in filament channels.

The study by \citet{yeates2010} used an alternative, magneto-frictional model with surface driving. This gives less accurate results for the detailed structure within active regions but is better able to reproduce the formation of filament channels and other non-potential structures outside them. A similar model will be used in the present paper and discussed further in Section \ref{s:setup}. Unlike \citet{chifu2022}, the study of \citet{yeates2010} covered Solar Cycle 23, looking at six different periods over the cycle. They found that the free magnetic energy increased by a factor of 15 between solar minimum and solar maximum, while the total magnetic energy increased by a factor of 8. The actual value of free energy during Maximum was around $2\times 10^{33}\,\mathrm{erg}$, a little below that found by \citet{chifu2022} for Solar Cycle 24.

The time-dependent magneto-frictional models also have the advantage of including the formation and eruption of magnetic flux ropes \citep{mackay2006, yeates2009}. \citet{yeates2010} defined flux ropes using the magnetic pressure and tension forces, then identified ejections by looking at the outward velocity. They found that the number of ejections increased from Minimum to Maximum by a factor of 8, proportional to the sunspot number as in the observations of coronal mass ejections. \citet{lowder2017} analysed a continuous magneto-frictional simulation from 1996 to 2014, with a new methodology for flux rope identification based on thresholding of field line helicity. Again, they found that the eruption rate followed solar activity. In addition, the model predicted a non-zero number of eruptions during the very weak Minimum between Cycles 23 and 24, which also seems to be consistent with observations. Both of these previous magneto-frictional models were driven by idealised bipolar active regions derived from US National Solar Observatory synoptic magnetograms.

Whilst the general trend of increasing non-potentiality from Solar Minimum to Solar Maximum is well established, the aim of this paper is to be more quantitative. We use a model that is able to approximate the evolution of non-potential structure on a global scale, rather than treating individual active regions in isolation. In particular, we update our earlier findings on solar cycle variation of non-potentiality during Cycle 23 to new magneto-frictional simulations of Cycle 24 based on HMI data. After describing the model and implementation (Section \ref{s:setup}), we will consider in Section \ref{s:refsim} a reference simulation where all active regions emerge untwisted. Given the presence of magnetic reconnection and outflow in our model, we would expect this to be a practical lower bound for the amount of free energy in the corona. In Section \ref{s:results}, we will then discuss the effect of emerging twisted active regions, before giving our conclusions in Section \ref{s:conclusions}.

This study benefits from two technical advances. The first is the emergence of active regions with observed shapes rather than idealised magnetic bipoles \citep{yeates2022}. The surface flux transport study by \citet{yeates2020} showed that in Cycle 24 the idealised bipoles lead to a 24\% overestimate of the overall axial dipole moment, but the effect on the non-potential coronal magnetic field has not yet been investigated. The second advance is the development of field line helicity as an additional tool for probing the non-potential structure of the coronal magnetic field \citep{berger1988, yeates2016,lowder2017, yeates2020a, moraitis2021}.

\section{Model setup}\label{s:setup} 

We simulate the corona using a magneto-frictional model for the mean (large-scale) magnetic field ${\bm B}$, coupled to an evolving surface flux transport model on the solar surface. This approach was introduced by \citet{vanballegooijen2000} and extended to the global corona by \citet{yeates2008}. It has since been applied in numerous studies, including investigations of filament formation \citep[e.g,][]{mackay2000, yeates2009a, mackay2018}, coronal mass ejections \citep[e.g.,][]{mackay2006, yeates2009, lowder2017,bhowmik2021}, sympathetic flares \citep{schrijver2013}, the middle corona \citep{meyer2020}, or the interplanetary magnetic field/solar wind \citep{yeates2010a, edwards2015}. It has also been applied to other stars \citep[e.g.,][]{lehmann2017}, by other researchers \citep{hoeksema2020}, and coupled to full magnetohydrodynamic simulations \citep[e.g.,][]{yardley2021, hayashi2021}. This experience has guided our choice of reference parameter values; the effect of varying these parameters is considered in Appendix \ref{s:params}.

\subsection{Magneto-frictional model}\label{s:mf} 

Our magneto-frictional model for the mean/large-scale magnetic field is based on the following main assumptions.
\begin{enumerate}
    \item The mean magnetic field ${\bm B}$ in the corona originates from the emergence of macroscopic active regions. In this paper we neglect small-scale ephemeral regions although they do contribute to the total magnetic energy in the real corona and potentially even to the magnetic helicity on large-scales, through an inverse cascade \citep{mackay2014}. Our treatment of emerging regions is described in Section \ref{s:ems}.
    \item The high magnetic Reynolds number implies a near-ideal evolution of ${\bm B}$, so it is important to retain the mean-field induction equation
    \begin{equation}
    \frac{\partial {\bm B}}{\partial t} = -\nabla\times{\bm E}, \qquad \bm{E} = - {\bm v}\times{\bm B} + {\bm N},
    \label{eqn:inducb}
    \end{equation}
    where the non-ideal term ${\bm N}$ is small. This means in particular that ${\bm B}$ holds a memory of previous interactions, allowing topological structure to build over time.
    \item The high Alfv\'en speed in the corona makes ${\bm B}$ respond rapidly to boundary motions on the solar surface (Section \ref{s:boundary}), and flux emergence from the solar interior (Section \ref{s:ems}). Accordingly, we neglect plasma forces and magnetohydrodynamic waves, and approximate the fluid momentum equation with an artificial velocity
    \begin{equation}
    {\bm v} = \frac{{\bm J}\times{\bm B}}{\nu B^2} + v_{\rm out}(r)\hat{\bm r}, \qquad {\bm J} = \nabla\times{\bm B}.
    \label{eqn:ufric}
    \end{equation}
    The first term represents the magneto-frictional assumption \citep{chodura1981,yang1986} and ensures relaxation towards a force-free equilibrium ${\bm J}\times{\bm B}={\bm 0}$,  albeit modified here by the second term. We also assume low plasma-beta and neglect  plasma pressure and gravity, although this is a simplification compared to reality \citep[cf.][]{chen2022}.
    \item In the outer corona, the solar wind starts to influence ${\bm B}$, preventing it from being force-free. This is accounted for by the radial outflow $v_{\rm out}(r)=v_w(r/R_1)^{11.5}$ in Equation \eqref{eqn:ufric}, which models the radially distending effect of the solar wind on closed-field arcades \citep{mackay2006}, and reduces the solution's sensitivity to the particular choice of outer boundary height, $R_1$ \citep{rice2021}. We set $R_1=2.5R_\odot$ and fix the outflow speed $v_w=100\,\mathrm{km}\,\mathrm{s}^{-1}$ in our reference model.
    \item The coronal ${\bm B}$ relaxes with respect to a rest frame rotating with the Sun. Our simulations use the Carrington frame.
\end{enumerate}

The non-ideal electric field, ${\bm N}$, in Equation \eqref{eqn:inducb} represents turbulent diffusion -- the net effect of unresolved small-scale fluctuations on the mean magnetic field. Here we assume the hyperdiffusive form
\begin{equation}
    {\bm N} = -\eta_h\frac{{\bm B}}{B^2}\nabla\cdot\left[B^2\nabla\left(\frac{{\bm J}\cdot\bm{B}}{B^2}\right)\right],
    \label{eqn:hyper}
\end{equation}
which prevents volume dissipation of relative helicity (with respect to a potential field) since the corresponding volume dissipation term $\int_V{\bm E}\cdot{\bm B}\,dV$ \citep{berger1984} becomes a boundary integral \citep{boozer1986, vanballegooijen2008}. We set the reference value $\eta_h=10^{11}\,\mathrm{km}^4\,\mathrm{s}^{-1}$ following \citet{vanballegooijen2007}. The recent study of \citet{mackay2022} suggests that hyperdiffusion gives quite a conservative estimate of the amount of free energy in the corona, compared to other possible forms for ${\bm N}$, although further observational comparison is needed to be sure of the most realistic approach.

The ``friction'' coefficient $\nu$ in equation \eqref{eqn:ufric} controls the speed of relaxation relative to hyperdiffusion and boundary driving. To reduce computational cost, we follow \citet{mackay2000} and reduce the relaxation rate near the poles, setting $\nu = \nu_0 (r\cos\lambda)^{-2}$, where $\lambda$ denotes latitude. In our reference model the amplitude is set to $\nu_0=2.8\times 10^{5}\,\mathrm{s}$.

In our numerical implementation, we write ${\bm B}=\nabla\times{\bm A}$ and solve the uncurled version of Equation \eqref{eqn:inducb} for the vector potential ${\bm A}$, using a finite volume method with a staggered mesh so as to conserve magnetic flux. The mesh has equally spaced points in $\log(r/R_\odot)$, sine latitude and longitude, with a resolution of $(60,180,360)$ cells. The equations are discretised using second-order ``mimetic'' spatial differences (discretising Stokes' Theorem), with a modified stencil at the polar grid points. Upwinding is used for the radial outflow term. Time-stepping uses a second-order Runge-Kutta method.
The simulations run from 12 June 2010 to 31 December 2019.

\subsection{Initial conditions}\label{s:init} 

To initialise the simulation we use a potential field source surface (PFSS) extrapolation taken from an imposed $B_r$ distribution on $r=R_\odot$, assuming ${\bm J}={\bm 0}$ in the corona with ${\bm B}\times\hat{\bm r}={\bm 0}$ on the outer source surface boundary, $r=R_1$. For the imposed $B_r$ distribution, we use the radial component, pole-filled HMI map for Carrington rotation CR 2097 from the \texttt{hmi.synoptic\_mr\_polfil\_720s} series \citep{sun2018}. To reduce its resolution to be comparable to our simulations of the mean field, this map is first smoothed by multiplying the spherical harmonic coefficients by a Gaussian filter $\mathrm{e}^{-c_wl(l+1)}$, with $c_w=5\times 10^{-4}$.

The initial potential field extrapolation is computed using the author's finite difference code \citep{yeates2018code, stansby2020}, which ensures that ${\bm J}={\bm 0}$ to machine precision for our particular discretisation scheme.

\subsection{Boundary conditions}\label{s:boundary} 

At the inner boundary the horizontal electric field from Equation \eqref{eqn:inducb} is replaced by
\begin{equation}
    {\bm E}_\perp\big|_{r=R_\odot} = {\bm E}_\perp^{\rm em} -{\bm v}_s\times\big(B_r\hat{\bm r}\big) + \eta_0\nabla\times\big(B_r\hat{\bm r}\big).
\end{equation}
The first term, ${\bm E}_\perp^{\rm em}$, represents emergence of new active regions and will be described in Section \ref{s:ems}.
The final two terms represent, respectively, advection by large-scale surface flows and supergranular diffusion (the net effect of small-scale surface motions). The large-scale flow velocity is assumed to take the steady form
${\bm v}_s = v_\theta(\theta)\hat{\bm\theta} + R_\odot\sin\theta\,\Omega(\theta)\hat{\bm\phi}$,
where $v_\theta$ denotes the meridional (poleward) flow and $\Omega(\theta)$ is the angular velocity of differential rotation. In this study we neglect any additional injection of magnetic helicity into the mean field from small-scale photospheric flows, although this could be included in parameterised form \citep[see][]{mackay2014}. 

The surface flux transport parameters ${\bm v}_s$ and $\eta_0$ were previously calibrated for the same emerging regions by \citet{yeates2020} to ensure a reasonable match to the observed time series of axial dipole strength. This led to $v_\theta(\theta) = D_u\cos\theta\sin^{p_0}\theta$, with $p_0=2.33$ and $D_u = 0.041\,\mathrm{km}\,\mathrm{s}^{-1}$ (giving a peak flow speed of $15\,\mathrm{m}\,\mathrm{s}^{-1}$), and $\eta_0 = 350\,\mathrm{km}^2\,\mathrm{s}^{-1}$. The calibration did not constrain the differential rotation, so in this paper we use the established angular velocity profile from \citet{snodgrass1990} of $\Omega(\theta) = 0.18 - 2.396\cos^2\theta - 1.787\cos^4\theta$ (degrees per day in the Carrington frame).

The resulting evolution of $B_r$ on the solar surface was illustrated in Figure 5 of \citet{yeates2020}. In that paper, it was also determined that if emerging regions are replaced by idealised BMRs, the meridional flow amplitude should be increased to $D_u=0.055 \,\mathrm{km}\,\mathrm{s}^{-1}$ (giving a peak flow speed of $20\,\mathrm{m}\,\mathrm{s}^{-1}$), to counter the spurious enhancement in end-of-cycle axial dipole strength resulting from the BMR approximation. This is adopted in our simulation run BMR0 below.

At both inner and outer boundaries, we impose ${\bm J}\times\hat{\bm r}={\bm 0}$, so as to avoid any flux of magnetic energy into or out of the domain due to the magneto-friction term in Equation \eqref{eqn:ufric}. To compute ${\bm N}$ on these boundaries, we impose zero-gradient conditions on the radial component of the hyperdiffusive flux $B^2\nabla({\bm J}\cdot{\bm B}/B^2)$.

\subsection{Emerging regions}\label{s:ems} 

\begin{figure}
    \centering
    \includegraphics[width=\textwidth]{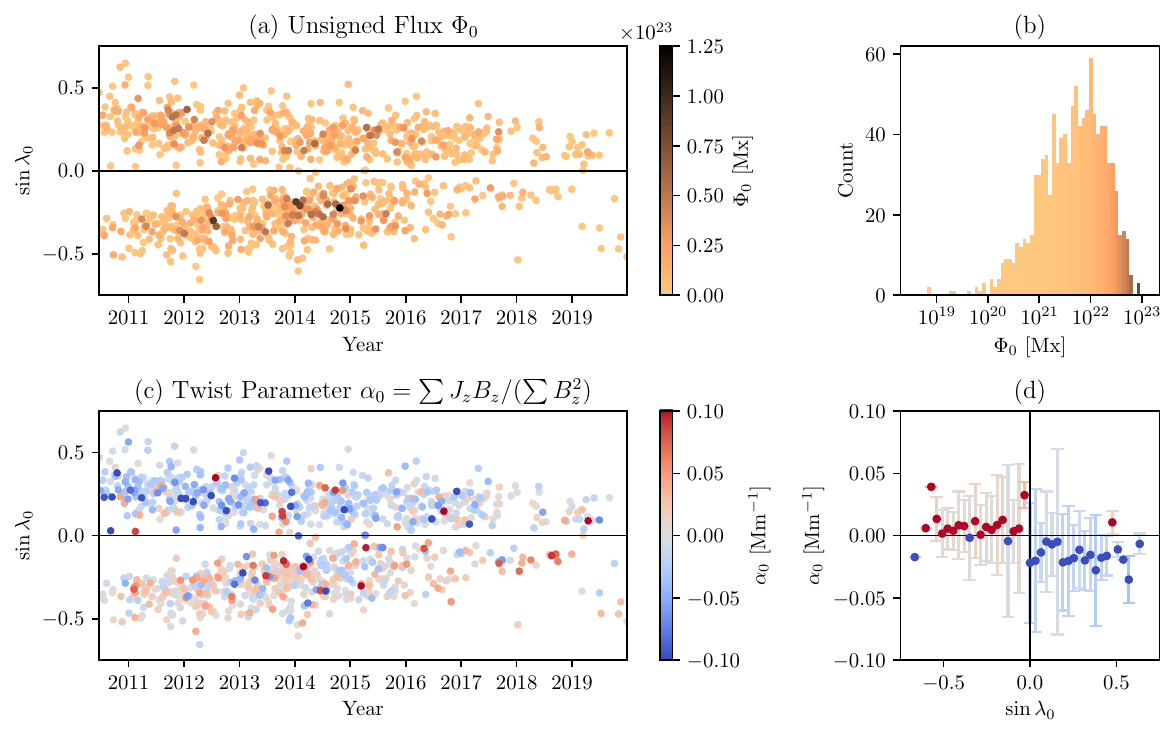}
    \caption{Emerging regions determined from the HMI/SHARPs data, shown in a time-latitude ``butterfly'' plot and coloured by either (a) unsigned flux, $\Phi_0$ or (c) observed twist, $\alpha_0$. Panel (b) shows a histogram of the $\Phi_0$ values, while (d) shows the mean and standard deviation of the $\alpha_0$ values in equal sine-latitude bins. 
    There is a hemispheric tendency in $\alpha_0$ with $\langle\alpha_0\rangle_{\lambda_0>0}=-0.0147\,\mathrm{Mm}^{-1}$ and $\langle\alpha_0\rangle_{\lambda_0<0}=0.00510\,\mathrm{Mm}^{-1}$. The percentage of regions obeying the hemispheric rule is $75.9\%$ in the North and $60.6\%$ in the South.}
    \label{fig:regions}
\end{figure}

We use the dataset of active regions extracted by \citet{yeates2020} from Spaceweather HMI Active Region Patch (SHARP) data \citep{bobra2014}. Briefly, a single line-of-sight magnetogram for each region is selected at the time closest to central meridian, and remapped to the simulation grid. The dataset is filtered to remove regions that are (i) too unbalanced in flux; (ii) too small to resolve; or (iii) repeat observations of the same region on an earlier disk passage. This leaves surface $B_r$ maps for 1072 regions during the simulation period. Figure \ref{fig:regions}(a) shows the time-latitude distribution of regions, coloured by their unsigned magnetic flux, $\Phi_0$ (including both polarities). The distribution of region fluxes is shown in Figure \ref{fig:regions}(b).

Figure \ref{fig:regions}(c) shows the same SHARP regions but coloured instead by the observed ``twist'' parameter $\alpha_0 = \sum J_zB_z/(\sum B_z^2)$ from the SHARP metadata \citep[\texttt{MEANALP} in Table 3 of][]{bobra2014}. Here the sum is taken over all observed pixels within the SHARP mask. This incorporates vector magnetogram information through $J_z$, and will be used here to set the helicity of emerging regions in two simulation runs. There is considerable scatter in the observed $\alpha_0$ values. Some of this is likely real -- resulting from the variation in helicity between different active regions \citep[e.g.,][]{georgoulis2009} -- but some is likely due to measurement error. In addition, it must be remembered that the $\alpha_0$ parameter does not correspond directly with the full current helicity, nor the magnetic helicity which is the underlying conserved quantity \citep[cf.][]{russell2019}. Nevertheless, the latitude distribution in Figure \ref{fig:regions}(d) shows that $\alpha_0$ recovers the expected hemispheric helicity trend of negative in the North and positive in the South \citep{pevtsov2003, liu2014}.

Our technique for emerging each region in the three-dimensional simulation was motivated, described, and illustrated in detail by \citet{yeates2022}. Briefly, for a fixed time interval $T_{\rm em}=24\,\mathrm{hr}$ we impose a steady (surface) electric field
\begin{equation}
    {\bm E}^{\rm em}_\perp(\theta,\phi) = {\bm E}^0_\perp(\theta,\phi) - \nabla\frac{\Psi(\theta,\phi)}{T_{\rm em}}.
    \label{eqn:eem}
\end{equation}
The first term ${\bm E}^0_\perp$ is the local inductive electric field, which generates the given $B_r$ distribution while producing a coronal magnetic field with minimum possible complexity (close to potential). This is termed ``local'' because we fix ${\bm E}^0_\perp={\bm 0}$ outside of the emerging region.

\begin{figure}
    \centering
    \includegraphics[width=\textwidth]{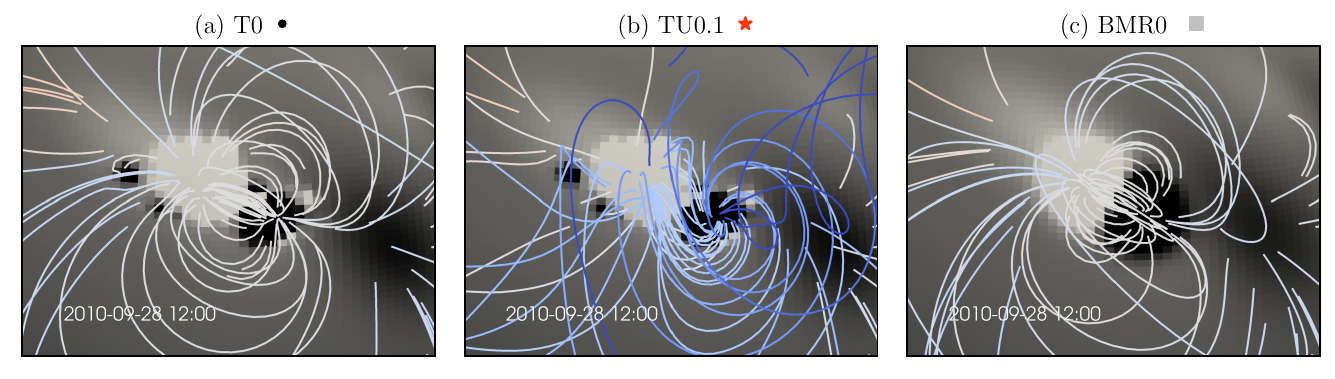}
    \caption{Three-dimensional renderings of an emerging region (SHARP 187/NOAA 11109 on 2010-09-28), immediately after completion of emergence in three of the simulation runs. The twist parameter is $\tau=0$ in (a,c) and $\tau=-0.1$ in (b). Grayscale pixels show $B_r$ on $r=R_\odot$ (white positive/black negative, saturated at $\pm 25\,\mathrm{G}$), while magnetic field lines are coloured by field line helicity (red positive/blue negative, saturated at $\pm 10^{22}\,\mathrm{Mx}$).}
    \label{fig:em-demo}
\end{figure}

The second term in \eqref{eqn:eem} does not change the final $B_r$ distribution but generates additional helicity in the three-dimensional field, allowing us to account for active regions that emerge in a non-potential state. In the reference run T0, this additional twisting is omitted, generating coronal active regions that are close to potential with low internal helicity. An example is shown in Figure \ref{fig:em-demo}(a).

In runs where additional twisting is included, we follow \citet{yeates2022} and set the twisting potential to
\begin{equation}
\Psi(\theta,\phi) = \tau b_0\langle f_{\rm pil}\rangle \overline{B}_r.
\label{eqn:twist}
\end{equation}
Here $\overline{B}_r$ is a smoothed magnetogram used to identify the polarity inversion lines, and $\langle f_{\rm pil}\rangle$ is a function which localises the twist near to them. The normalisation factor $b_0$ is chosen to ensure that the horizontal magnetic field generated by  twisting is $\approx|\tau B_r|$. The single dimensionless ``twist'' parameter $\tau$ then controls the helicity injection within the emerging region, with $\tau >0$ leading to positive helicity and $\tau <0$ to negative helicity. Figure \ref{fig:em-demo}(b) shows the same region as Figure \ref{fig:em-demo}(a), but emerged with $\tau=-0.1$ giving it negative helicity -- a reverse-S shaped twist.

Since the optimum value of $\tau$ to model each active region is unknown, we follow \citet{yeates2022} and vary $\tau$ in a parameter study. The simulation runs are summarised in Table \ref{tab:runs}. In TU0.05 and TU0.1, we simply fix a uniform $\tau$ value for all regions in each hemisphere. In some sense this maximises the energisation since it minimises cancellation between positive and negative helicity during the relaxation. The largest value of $|\tau|=0.1$ is chosen because larger values lead to unrealistically twisted coronal field lines \citep{yeates2022}. In order to consider the effect of varying helicity between active regions, runs TOb5 and TOb10 use different values of $\tau$ for each emerging region, chosen proportional to the observed twist values $\alpha_0$. Since the dimensions of $\tau$ and $\alpha_0$ are different, the proportionality constant $\ell$ in Table \ref{tab:runs} has dimensions of length. The values of $\ell$ are chosen so that the two runs have a comparable overall helicity to the two runs with uniform $\tau$.

Finally, for most runs, we use the observed SHARP active region shapes. For run BMR0, the regions are replaced by equivalent, idealised BMRs (bipolar magnetic regions), according to the prescription in \citet{yeates2020}. We take these to be untwisted, as illustrated in Figure \ref{fig:em-demo}(c).

\begin{table}
\caption{ Model runs with different emerging regions. Coronal parameters take values in all of these runs, namely $\nu_0=2.8\times 10^{5}\,\mathrm{s}$, $\eta_h=10^{11}\,\mathrm{km}^4\,\mathrm{s}^{-1}$, $v_w=100\,\mathrm{km}\,\mathrm{s}^{-1}$.
}
\label{tab:runs}
\begin{tabular}{llccccccccc}     
  \hline                   
Run  & Emerging Regions & Twist & \multicolumn{3}{c}{SFT Parameters}\\
     &  & $\tau$ & $\eta_0$ [$\mathrm{km}^2\,\mathrm{s}^{-1}$]& $D_u$ [$\mathrm{km}\,\mathrm{s}^{-1}$] & $p_0$ & \\
  \hline
T0  & Observed & 0.00 & 350 & 0.041 & 2.33\\
TU0.05 & Observed & $\pm0.05$ & 350 & 0.041 & 2.33\\
TU0.1 & Observed & $\pm0.10$ & 350 & 0.041 & 2.33\\
TOb5 & Observed & $\ell\alpha_0$, $\ell=5\,\mathrm{Mm}$ & 350 & 0.041 & 2.33\\
TOb10 & Observed & $\ell\alpha_0$, $\ell=10\,\mathrm{Mm}$ & 350 & 0.041 & 2.33\\
BMR0 & Idealised BMRs & 0.00 & 350 & 0.055 & 2.33\\
  \hline
\end{tabular}
\end{table}

\subsection{Cycle variation of the photospheric magnetic field}\label{s:phot}

\begin{figure}
    \centering
    \includegraphics[width=\textwidth]{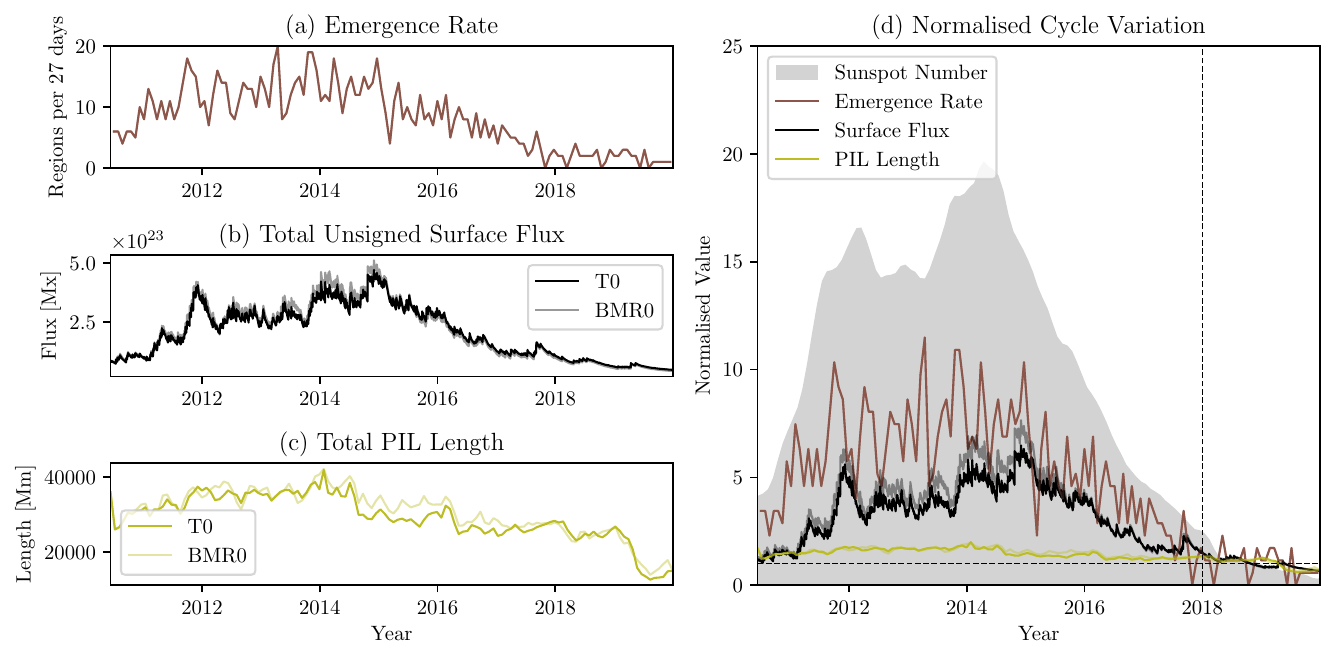}
    \caption{Cycle variation of the solar surface magnetic field in runs T0 (original region shapes) and BMR0 (idealised bipoles). Panels (a-c) show the quantities with their original units, while panel (d) shows all curves normalised to their mean value during 2018 and 2019 (to right of the vertical dashed line). Grey shading shows the 13-month smoothed monthly International Sunspot Number, normalised in the same way. (Source: WDC-SILSO, Royal Observatory of Belgium, Brussels.)}
    \label{fig:cycle-phot}
\end{figure}

The distribution of $B_r$ on the solar surface is unaffected by the choice of coronal parameters or emerging region twist. The longitude-averaged $B_r$ was shown as a function of time and latitude in Figure 5 of \citet{yeates2020}. Here, in Figures  \ref{fig:cycle-phot}(a-c), we show the emergence rate (number of emerging regions in 27-day bins), total unsigned magnetic flux,
\begin{equation}
    \Phi_0 = \oint_{r=R_\odot}|B_r|\,\mathrm{d}S,
\end{equation}
and total polarity inversion line (PIL) length. The darker curves show the reference run T0 (or equivalently TU0.05, TU0.1, TOb5, or TOb10). The feinter curves show run BMR0. This has the same emergence rate. However, whilst the BMR approximation does not change $\Phi_0$ for individual regions, it does slightly increase $\Phi_0$ around Solar Maximum because of differences in flux cancellation between regions, and the modified $B_r$ pattern also slightly increases the PIL length.

Figure \ref{fig:cycle-phot}(d) shows the same three quantities but normalised to their mean values during 2018 and 2019 (to the right of the vertical dashed line). This period is chosen to represent Solar Minimum rather than the start of the simulation because the HMI dataset driving the simulation does not extend all the way back to Solar Minimum. Similar to \cite{yeates2010}, we observe that (i) $\Phi_0$ is about 4-5 times higher at Maximum than Minimum, and (ii) the emergence rate in the simulation is about 7-8 times higher at Maximum than Minimum.
Note that the increase in observed sunspot number is rather higher than in the previous study, perhaps 20 times more at Maximum than Minimum. This is likely because the SHARP regions can contain multiple spots.

At the resolution considered, the PIL length shows a much more modest increase of less than a factor 2 from Minimum to Maximum, although the length fluctuates substantially over 2018 and 2019. This modest increase is again in line with \cite{yeates2010}.

\section{Reference simulation}\label{s:refsim} 

\begin{figure}
    \centering
    \includegraphics[width=\textwidth]{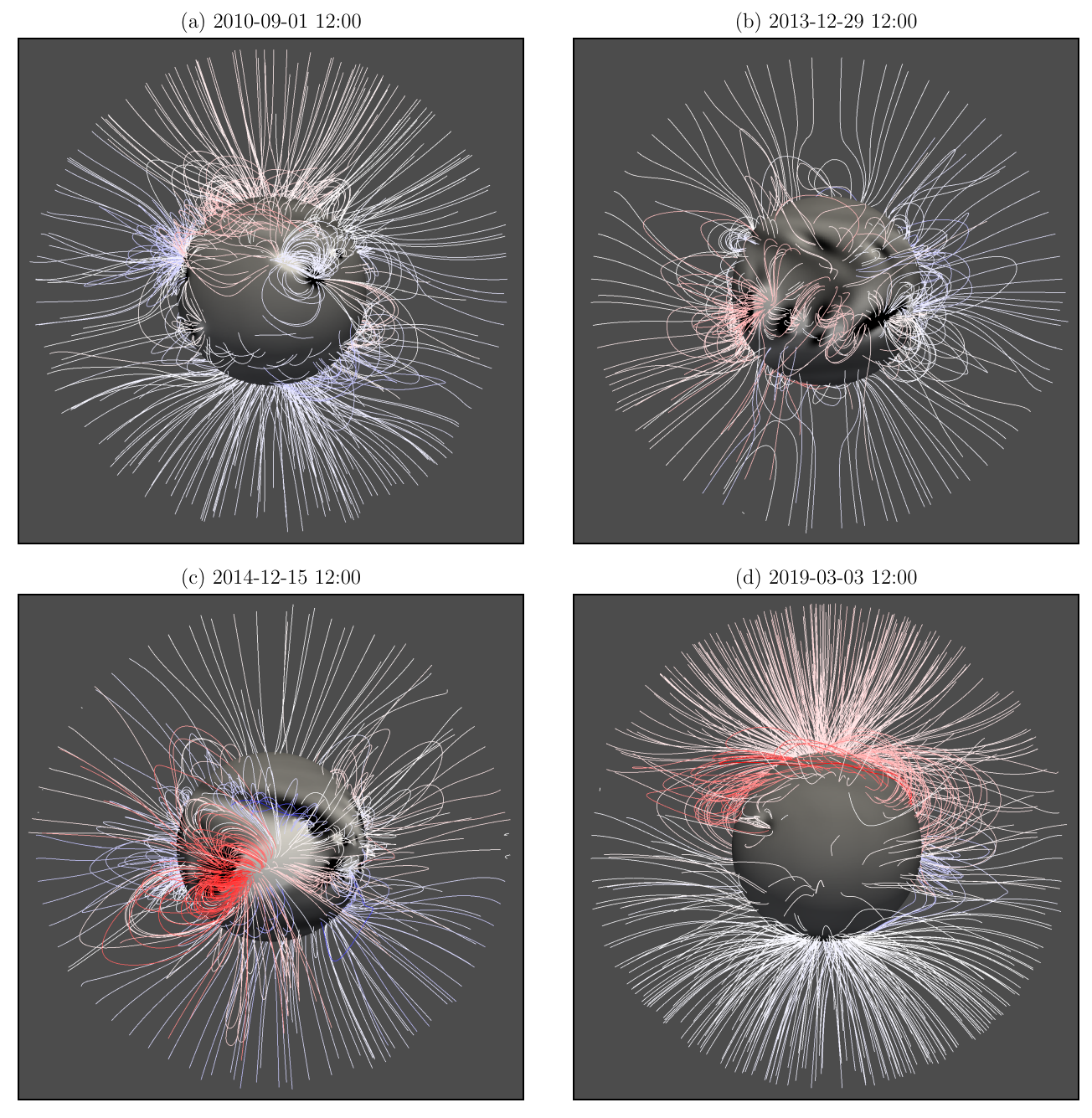}
    \caption{Four snapshots from the reference simulation, run T0, viewed from the direction of Earth at the corresponding time. Magnetic field lines are coloured by their field line helicity (Section \ref{s:nonpot}): red positive/blue negative, saturated at $\pm 10^{22}\,\mathrm{Mx}$. Gray shading shows $B_r$ on the solar surface, black negative/white positive, saturated at $\pm 15\,\mathrm{G}$.}
    \label{fig:corona}
\end{figure}

In this section we analyse the solar cycle variation of the reference simulation: run T0 in Table \ref{tab:runs}. This represents a conservative model where all active regions emerge with no internal twist/helicity. As evidenced by the selected snapshots in Figure \ref{fig:corona}, this still leads to a corona that is non-potential with concentrations of magnetic helicity in sheared arcades or twisted flux ropes \citep[cf.][]{yeates2016,bhowmik2021}. The associated free energy can build up at active latitudes, as in Figure \ref{fig:corona}(c), but it can also be stored in closed-field arcades at higher latitudes, as seen on the Northern polar crown in Figure \ref{fig:corona}(d). In the following subsections we will consider how quantitative properties of the simulated magnetic field vary between Solar Minimum and Solar Maximum.

\subsection{Energy}\label{s:en}

\begin{figure}
    \centering
    \includegraphics[width=\textwidth]{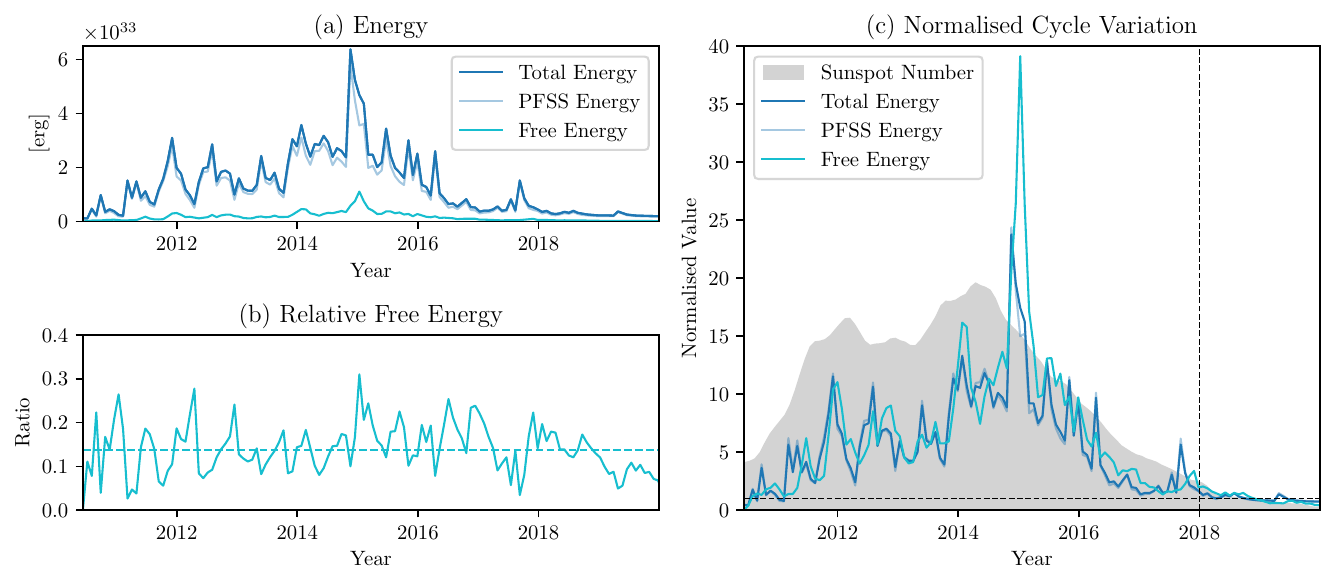}
    \caption{Cycle variation of the coronal magnetic energy in run T0, computed at 27-day intervals. Dark blue curves show the total energy, $E$, blue-grey the energy of a corresponding PFSS extrapolation, $E_{\rm p}$, and cyan the free energy, $E_{\rm free}$. Panel (a) shows the original values while (b) shows the relative free energy, $E_{\rm free}/E_{\rm p}$. Panel (c) shows the same curves as (a) but normalised as in Figure \ref{fig:cycle-phot} and compared to the observed sunspot number.}
    \label{fig:cycle-energy}
\end{figure}

Figure \ref{fig:cycle-energy}(a) shows total magnetic energy,
\begin{equation}
E = \int_V\frac{B^2}{8\pi}\,\mathrm{d}V,    
\end{equation}
along with the energy of a corresponding PFSS extrapolation. The cyan curve shows free energy,
\begin{equation}
E_{\rm free} = E - E_{\rm p}, \qquad E_{\rm p}= \int_V\frac{(B_{\rm p})^2}{8\pi}\,\mathrm{d}V,
\end{equation}
where ${\bm B}_{\rm p}$ is the current-free magnetic field in the coronal volume matching $B_{{\rm p}r}=B_r$ on both $r=R_\odot$ and $r=R_1$. Whilst this reference potential field technically differs from the PFSS field shown in Figure \ref{fig:cycle-energy}(a), their energies are indistinguishable on the scale of the plot. The peak around late 2014/early 2015 arises from a particularly strong activity complex -- the ``Great Solar Active Region'' NOAA 12192 \citep{sun2015} which was the largest since 1990. This already causes a substantial peak of magnetic energy in the PFSS model, and coincides with the peak of solar activity as noted by \citet{chifu2022}.

To measure the overall energisation, Figure \ref{fig:cycle-energy}(b) shows relative free energy, $E_{\rm free}/E_{\rm p}$. This fluctuates from one snapshot to the next, but on average is about $13.8\%$ over much of the cycle.
This is rather lower than many of the non-potential models compared by \citet{yeates2018}, where the ratio $E_{\rm free}/E_{\rm p}$ was in the range $40-50\%$. However, we will see in Section \ref{s:results} that $E_{\rm free}$ is substantially increased if the active regions emerge twisted.

Figure \ref{fig:cycle-energy}(c) shows how the three energies vary over the solar cycle, compared to the 2018-2019 period. Since energy is quadratic in ${\bm B}$, by comparing to the variation of photospheric flux in Figure \ref{fig:cycle-phot}(d) one would expect roughly a factor $16-25$ increase from Minimum to Maximum, and this is indeed the order of magnitude found. Note that $E$ and $E_{\rm p}$ show a similar relative increase, as does $E_{\rm free}$ albeit a little higher. However, one has to bear in mind that these are snapshots rather than running means; in particular, since $E_{\rm free}$ tends to lag $E_{\rm p}$ (because of the time taken for differential rotation to shear the field), the two energies -- which are fluctuating on a faster timescale -- are not necessarily sampled in phase. Thus we should not read too much into the precise heights of the peaks.

We note that \citet{yeates2010} found $E_{\rm free}$ to increase rather more from Minimum to Maximum compared to $E$, but here we find it much closer. However, this is again an effect of assuming the emerging regions to be untwisted, as we will see in Section \ref{s:results}.

\subsection{Current and helicity}\label{s:nonpot}

\begin{figure}
    \centering
    \includegraphics[width=\textwidth]{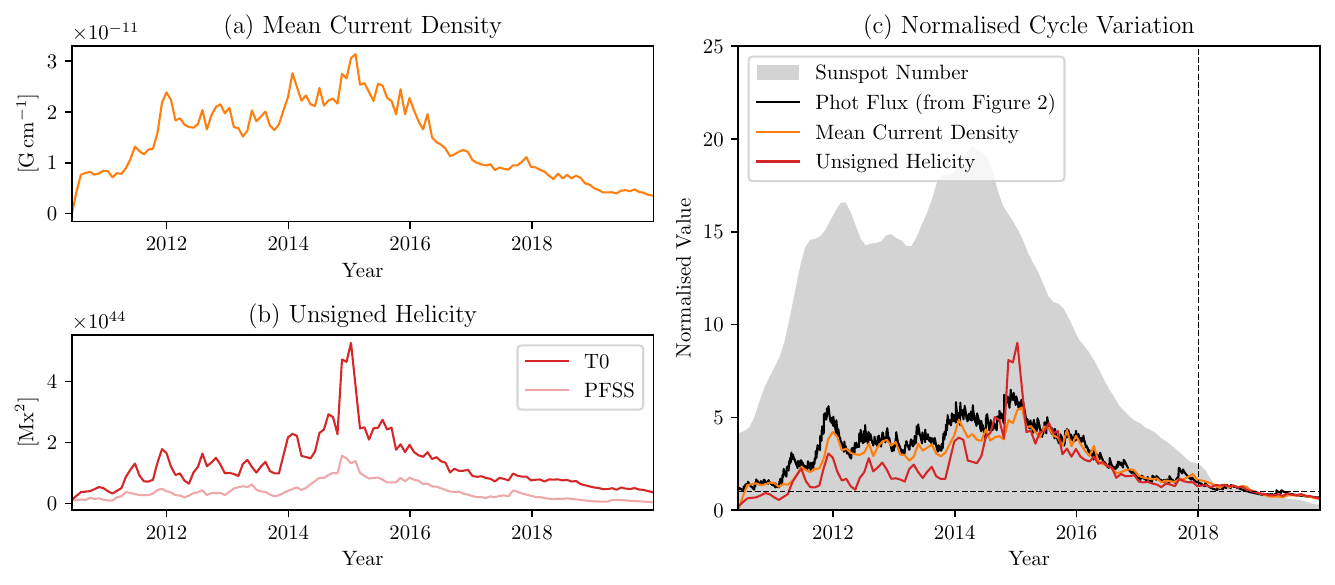}
    \caption{Cycle variation of (a) the mean current density $\langle J\rangle$ and (b) the unsigned helicity $\overline{H}$, in run T0, computed at 27-day intervals. Panel (c) shows the same curves as (a) and (b) but normalised as in Figure \ref{fig:cycle-phot} and compared to the observed sunspot number. The feint line in (b) shows $\overline{H}$ for corresponding PFSS extrapolations from the same surface $B_r$ distribution.}
    \label{fig:cycle-nonpot}
\end{figure}

Free energy is not the only way to characterise the complexity of the coronal magnetic field. Figure \ref{fig:cycle-nonpot} shows two further measures: mean current density and  unsigned helicity.

Mean current density -- shown in Figure \ref{fig:cycle-nonpot}(a) is defined as
\begin{equation}
    \langle{J\rangle} = \frac{\int_V|\nabla\times{\bm B}|\,\mathrm{d}V}{\int_V\,\mathrm{d}V},
\end{equation}
where $V$ is the full simulation volume. Figure \ref{fig:cycle-nonpot}(c) shows that it follows a similar 4-5 fold increase to the magnetic flux from Figure \ref{fig:cycle-phot}.

Unsigned helicity -- shown in Figure \ref{fig:cycle-nonpot}(b) -- is a measure of the topological complexity of the coronal magnetic field. It is defined as
\begin{equation}
    \overline{H} = \frac12\int_{\partial V}\big|\mathcal{A}B_r\big|\,\mathrm{d}S,
    \label{eqn:hbar}
\end{equation}
where $\mathcal{A}$ is the field line helicity 
\begin{equation}
    \mathcal{A}(L) = \int_L{\bm A}^*\cdot\,\mathrm{d}{\bm l},
    \label{eqn:flh}
\end{equation}
for a magnetic field line $L$ traced from a given point on $\partial V$, and ${\bm A}^*$ is an appropriate vector potential. Figure \ref{fig:cycle-nonpot}(c) shows a similar rate of increase in $\overline{H}$ from Minimum to Maximum compared to $\langle J\rangle$, albeit a rather different time profile, relatively lower in the increasing phase of the cycle. Like magnetic energy,  $\overline{H}$ also shows a peak in late 2014. This peak is present even in the PFSS model \citep[as explained by][]{yeates2020a}, but is significantly enhanced in the non-potential model -- even in run T0 -- because the strong active region acts a seed for shearing by differential rotation.

Our rationale for using $\overline{H}$ rather than the simpler-to-compute $\int_V|{\bm A}\cdot{\bm B}|\,\mathrm{d}V$ is that $\overline{H}$ is a topological invariant: it is unchanged under ideal deformations within the domain that fix the field line footpoints on the boundary, whereas the latter is not (and neither is $\langle J\rangle$). Thus it is a more robust measure than $\int_V|{\bm A}\cdot{\bm B}|\,\mathrm{d}V$. Our rationale for using $\overline{H}$ rather than the commonly-used relative helicity invariant \citep{berger1984} is the fact that relative helicity is a signed quantity. It would be a poor measure of topology for our global model because it would incorporate cancellation between regions of positive and negative helicity, even if these regions are remotely located on the Sun and do not interact in reality. This departure from what has become routine practice in solar physics is further discussed in Appendix \ref{s:hr}.

The specific values of $\mathcal{A}(L)$ and hence of $\overline{H}$ depend on the chosen gauge of ${\bm A}^*$. However, recent work suggests that there is a meaningful ``canonical'' choice of gauge \citep{berger2018, yeates2020a, xiao2023}, having the form ${\bm A}^*=\nabla\times(P\hat{\bm r}) + T\hat{\bm r}$ for suitable $P$ and $T$. Since ${\bm A}^*$ differs from the gauge of ${\bm A}$ used to solve Equation \eqref{eqn:inducb}, we compute ${\bm A}^*$ numerically to find $\mathcal{A}$ and $\overline{H}$. Whilst not unique, using this gauge consistently for all of our calculations allows us to compare different time snapshots and different runs. Moreover, as we will see in Section \ref{s:results}, this gauge choice leads to a meaningful correlation between $\overline{H}$ and free energy.

Note that even potential fields can have non-zero $\overline{H}$ if their boundary $B_r$ distribution lacks axisymmetry -- this behaviour was demonstrated for typical solar-like configurations in \citet{yeates2020a}. For run T0, the feint curve in Figure \ref{fig:cycle-nonpot}(b) shows that $\overline{H}$ in the PFSS model is approximately $33\%$ of $\overline{H}$, following a similar cycle trend.
Correlation between them arises because sub-structure in the PFSS field acts as a seed for subsequent shearing by differential rotation, which amplifies $\overline{H}$ in run T0. However, $\overline{H}$ also fluctuates more rapidly in the magneto-frictional simulation due to sudden reconfigurations such as flux rope ejections.

\subsection{Eruptivity}\label{s:eruptions}

In this paper we avoid identifying individual magnetic flux ropes, as this is a significant task in itself and the statistics would depend on the method used and definition of a flux rope \citep[but for two possible approaches see][]{yeates2009, lowder2017}. Instead, we have identified a proxy for the eruption rate that can be computed without the need to identify individual structures or track them over time. Importantly, this proxy can be computed from global diagnostics which can be saved at high time cadence, rather than requiring three-dimensional magnetic snapshots.

\begin{figure}
    \centering
    \includegraphics[width=\textwidth]{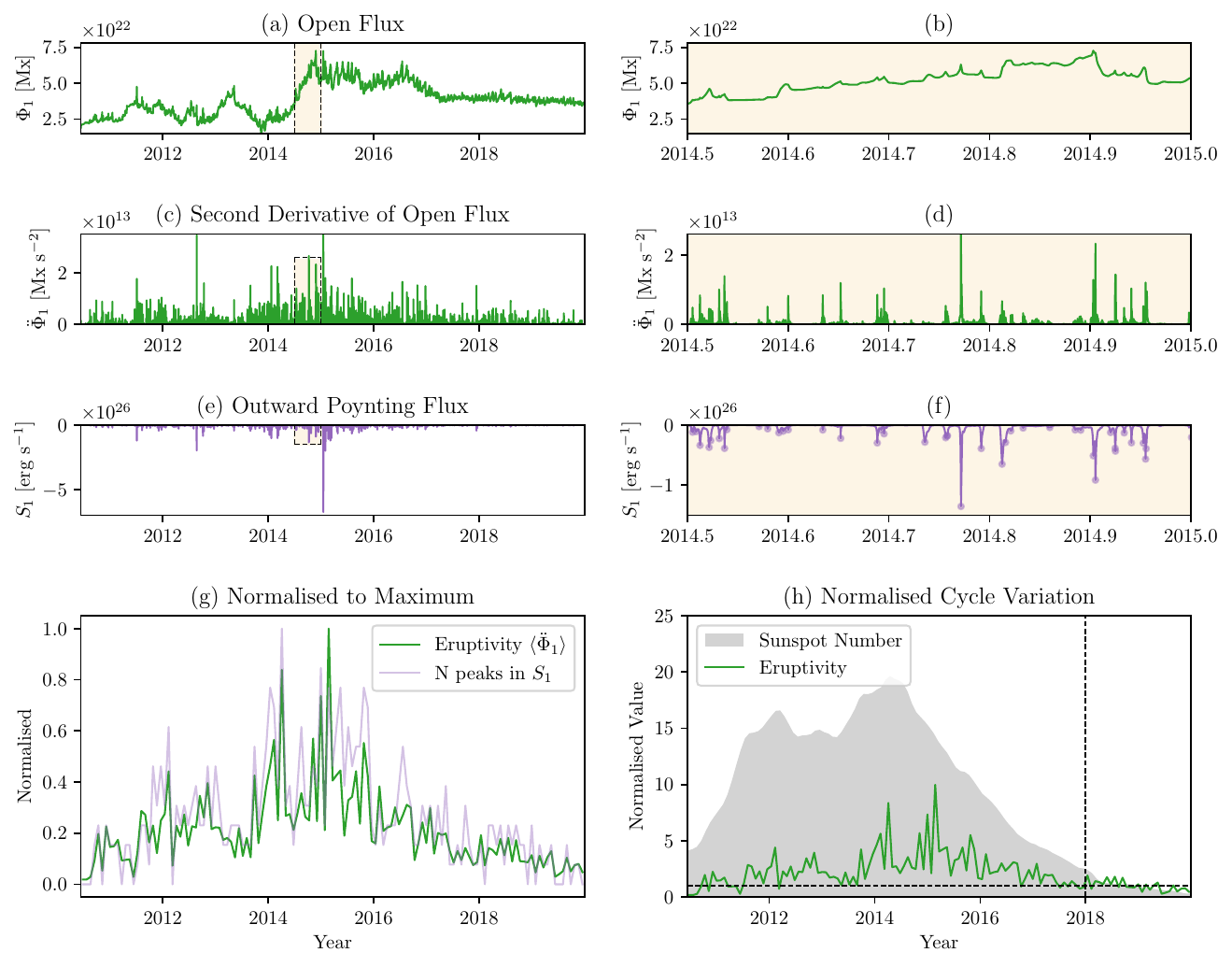}
    \caption{Time series used to compute the eruptivity proxy, $\langle\ddot{\Phi}_1\rangle$, shown for run T0. Panel (a) shows the open magnetic flux, $\Phi_1$ over the full simulation, with enlargement of a selected period in (b). Panels (c-d) show the second time derivative, $\ddot{\Phi}_1$. For comparison, panels (e-f) show the Poynting flux, $S_1$, through the outer boundary, with dots in (f) indicating peaks. Panel (g) shows the 27-day mean $\langle\ddot{\Phi}_1\rangle$, overlayed with the number of $S_1$ peaks in 27-day bins, both normalised to their maximum values. Panel (h) shows $\langle\ddot{\Phi}_1\rangle$ normalised as in Figure \ref{fig:cycle-phot} and compared to the observed sunspot number.}
    \label{fig:eruptivity}
\end{figure}

Our chosen eruptivity proxy is the second time derivative of unsigned open magnetic flux,
\begin{equation}
    \Phi_1 = \oint_{r=R_1}|B_r|\,\mathrm{d}S.
\end{equation}
As illustrated by \cite{bhowmik2021}, eruptions of individual structures in the model lead to temporary enhancements in $\Phi_1$. Taking the second time derivative, $\ddot\Phi_1$, removes the underlying secular variation and extracts these peaks. The evolution of this quantity is illustrated for run T0 in Figures \ref{fig:eruptivity}(a-d).

Another signature of eruptions are short ``bursts'' in the Poynting flux of magnetic energy through the outer boundary. From Equations \eqref{eqn:inducb}-\eqref{eqn:hyper}, and our boundary conditions, this takes the form
\begin{equation}
S_1 = -\frac{1}{4\pi}\oint_{r=R_1}{\bm E}\times{\bm B}\cdot\hat{\bm r}\,dS = -\frac{v_{\rm out}(R_1)}{4\pi}\oint_{r=R_1}|{\bm B}\times\hat{\bm r}|^2\,dS.
\end{equation}
Due to the radial outflow, the energy flux is purely outward (negative). Moreover, it depends only on the horizontal magnetic field, so is negligible in equilibrium, spiking only during eruptive events. It is shown for run T0 in Figures \ref{fig:eruptivity}(e-f). Comparing the enlarged Figures \ref{fig:eruptivity}(d) and (e) shows that the timings of peaks in $S_1$ (labelled with small circles) correspond well to peaks in $\ddot\Phi_1$.

For clearer comparison over the full solar cycle, we take our eruptivity proxy to be the 27-day average, $\langle\ddot\Phi_1\rangle$. This is shown by the green curve in Figure \ref{fig:eruptivity}(g). The purple curve shows the number of peaks in $S_1$ in 27-day bins, 
which is seen to follow a similar variation over the cycle. The peaks were counted requiring a  mininium-to-maximum height of at least the mean value $2\times 10^{24}\,\mathrm{erg}\,\mathrm{s}^{-1}$, but varying this threshold changes only the overall normalisation in Figure \ref{fig:eruptivity}(g) and not the relative variation over the cycle. Thus $\langle\ddot\Phi_1\rangle$ gives a meaningful proxy for the normalised eruption rate. Note that $S_1$ itself would peak more strongly at Maximum because it is quadratic in ${\bf B}$. Our chosen measure instead counts the rate of eruptions, with less regard to size. This is simpler to compare with observations.

For run T0, Figure \ref{fig:eruptivity}(h) shows the cycle variation of $\langle\ddot\Phi_1\rangle$ when normalised by its values in 2018 and 2019, as for previous quantities. We observe an increase from Minimum to Maximum by a factor comparable with $\Phi_0$, $\langle J\rangle$, or $\overline{H}$, but rather less than $E_{\rm free}$. This contrasts with \citet{yeates2010} where the rate of eruptions was found to increase more substantially over the cycle, more akin to the sunspot number or total energy. Furthermore, $\langle\ddot\Phi_1\rangle$ does not follow the shape of the sunspot curve, peaking in late 2014 but weaker before 2014. In particular, the eruptions themselves do not correlate with the timing of individual active region emergences -- some are caused more-or-less directly by emergence, but others arise from the longer term shearing of remnant flux, as in previous magneto-frictional models \citep[e.g.][]{mackay2006}. However, we will see in the next section that the eruptivity is also affected by the helicity of emerging regions.

\section{Effect of emerging region properties}\label{s:results}

We have repeated the analysis of coronal properties for each simulation run in Table \ref{tab:runs}.

\subsection{Emerging region twist -- cycle variation}

To investigate the effect of injecting helicity in the emerging regions, we compare runs TU0.05, TU0.1, TOb5 and TOb10 to the reference run T0. All of these runs share the same evolution of $B_r$ on the solar surface as run T0, with the only difference being the non-zero twist parameters $\tau$.

\begin{figure}
    \centering
    \includegraphics[width=\textwidth]{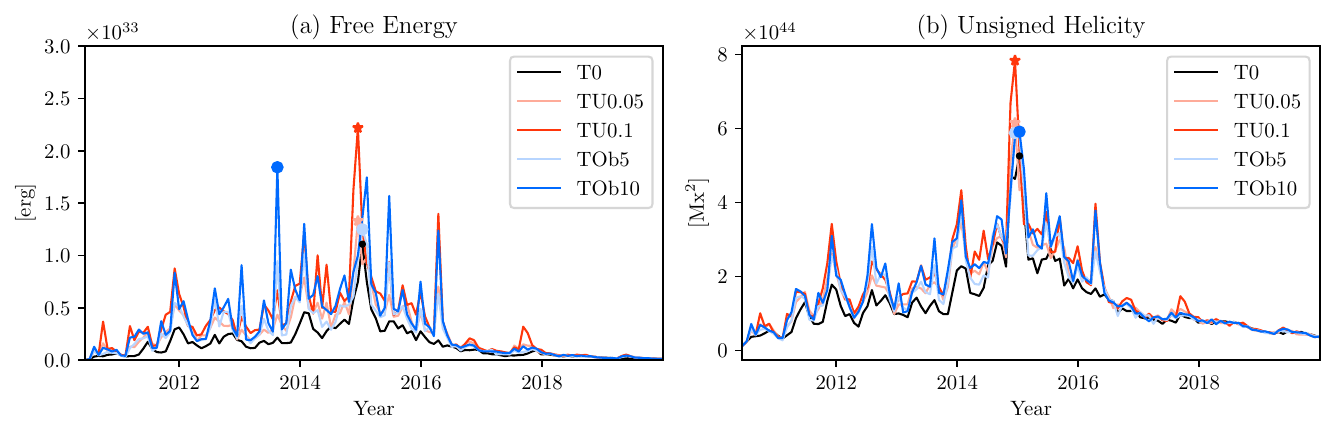}
    \caption{Time evolution of (a) $E_{\rm free}$ and (b) $\overline{H}$, in the coronal simulations with different emerging region properties. Here colours distinguish the different runs, with symbols indicating the peaks. The curves for run T0 are repeated from Figures \ref{fig:cycle-energy}(a) and \ref{fig:cycle-nonpot}(b).}
    \label{fig:compare-energy}
\end{figure}

Figure \ref{fig:compare-energy} compares the time evolution of $E_{\rm free}$ and 
 $\overline{H}$ for the different runs. Both quantities are enhanced in all of the runs with $\tau\neq 0$, compared to T0. Again, this is clearly related to periods of active region emergence, as one would expect given that the additional free energy is coming from the twisting of emerging regions. Figure \ref{fig:compare-energy}(a) for $E_{\rm free}$ can be compared with the NLFFF extrapolations in Figure 8 of \citet{chifu2022}. Even in run TOb10, our $E_{\rm free}$ looks to be perhaps $50\%$ lower on average, likely due at least in part to the lower resolution -- \citet{chifu2022} use (180,280,720) cells for analysis, and their $E_{\rm free}$ peaks at about $2.5\times 10^{33}\,\mathrm{erg}$ in late 2014. On the other hand we do see a similar overall pattern of short-term enhancements, with most of our models peaking similarly in late 2014. Our $E_{\rm free}$ shows sharper fluctuations, during which the magnitude can be comparable to \citet{chifu2022}. On the other hand, our $E_{\rm free}$ values around March 2015 look broadly consistent with those obtained by 
\citet{mackay2022} in their magneto-frictional runs with hyperdiffusion.
 
\begin{figure}
    \centering
    \includegraphics[width=\textwidth]{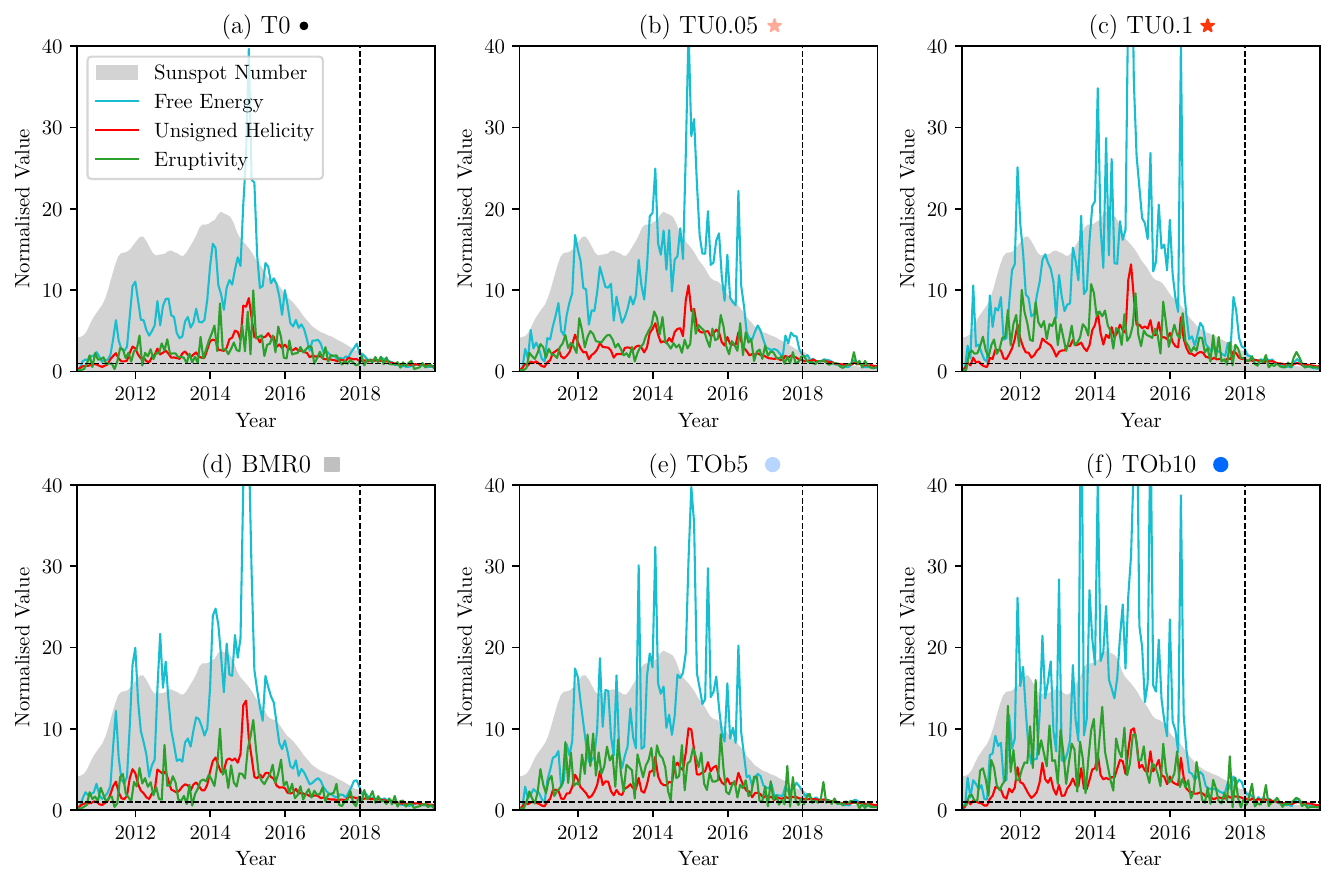}
    \caption{Cycle variation of parameters in the coronal simulations with different emerging region properties. Each panel corresponds to a different simulation run (Table \ref{tab:runs}), with all curves normalised as in Figure \ref{fig:cycle-phot} and compared to the observed sunspot number.}
    \label{fig:cycle-twist}
\end{figure}

Figures \ref{fig:cycle-twist}(b,c,e,f) show the normalised cycle variation of $E_{\rm free}$ and $\overline{H}$ for each run, as well as the eruptivity $\langle\ddot\Phi_1\rangle$. Again we see that, as we increase $|\tau|$ -- either the uniform value in runs TU0.05 and TU0.1 or the factor multiplying $\alpha_0$ for each region in runs TOb5 and TOb10 -- there is a greater contrast in non-potentiality of the corona from Minimum to Maximum. This is particularly seen in $E_{\rm free}$, which now shows a greater increase than the sunspot number. The pattern of variation in $\langle\ddot\Phi_1\rangle$ is also brought more into line with that of the sunspot number, in particular being enhanced during the rising phase of the cycle. Both of these findings bring the results into line with \citet{yeates2010}, removing the discrepancies seen in run T0.

\subsection{Emerging region twist -- cycle averages}

\begin{figure}
    \centering
    \includegraphics[width=\textwidth]{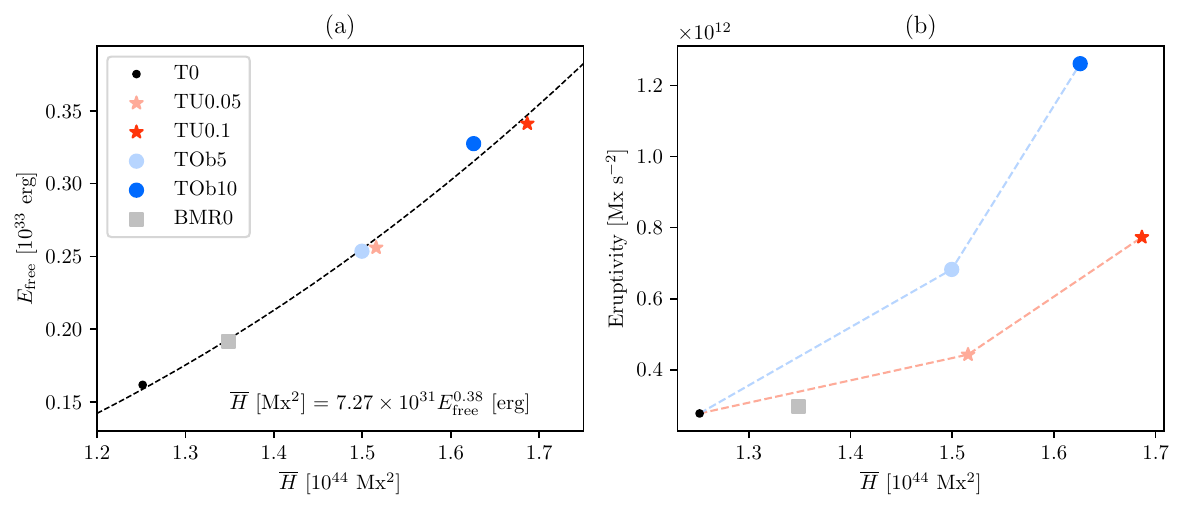}
    \caption{Relation between cycle-averaged unsigned helicity, $\overline{H}$ and (a) free energy, $E_{\rm free}$; or (b) eruptivity, $\langle\ddot\Phi_1\rangle$. Each symbol denotes a run with different emerging region properties. In (a) the dashed line shows a power-law fit with the given coefficients (in the indicated units), while in (b) separate dashed lines guide the eye between runs with uniform $\tau$ (red) and $\tau\propto\alpha_0$ (blue).}
    \label{fig:scatter}
\end{figure}

Figure \ref{fig:scatter} shows some interesting scatter plots for the simulations in Table \ref{tab:runs}, namely cycle-averaged free energy and eruptivity against cycle-averaged $\overline{H}$.
In Figure \ref{fig:scatter}(a), we see a clear power-law relation between the cycle-averages of $E_{\rm free}$ and $\overline{H}$, which holds for all of the runs (including BMR0). It is also clear how both quantities increase as the emerging region twist $\tau$ is increased.  The intermediate runs TU0.05 and TOb5 are roughly comparable to one another in both $\overline{H}$ and $E_{\rm free}$. Compared to the reference run T0, these two runs have approximately $20\%$ more unsigned helicity and $55\%$ more free energy. 

A similar relation between free energy and magnetic helicity has been found for data-driven force-free models of solar active regions \citep{tziotziou2012}. Those authors found a fit of $|H_R|=1.37\times 10^{14}\, E_{\rm free}^{0.897}$, where $E_{\rm free}$ is in ergs and $H_R$ is the usual relative magnetic helicity
 in $\mathrm{Mx}^2$. If this is extrapolated to a global value of $H_R=1.5\times 10^{44}\,\mathrm{Mx}^2$ -- approximately the cycle mean of $\overline{H}$ for run TU0.05 -- it predicts $E_{\rm free} \approx 3\times 10^{33}\,\mathrm{erg}$, which is an order of magnitude greater than our simulations. One possible explanation for the difference is the fact that $|H_R|\leq \overline{H}$, with equality only if $\mathcal{A}$ is uniform in sign throughout the region. Another possibility is that active regions have free energy further above the helicity threshold than the corona does as a whole.

 \begin{figure}
    \centering
    \includegraphics[width=\textwidth]{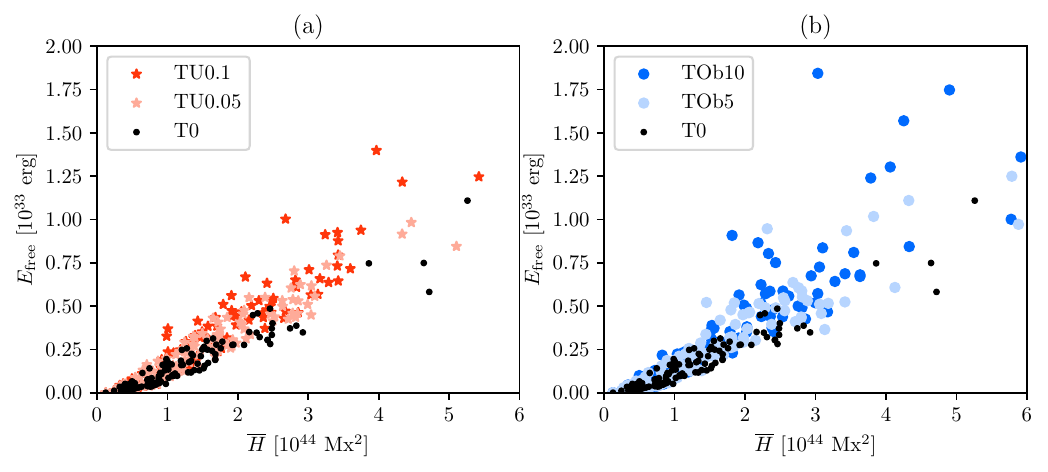}
    \caption{Relation between unsigned helicity, $\overline{H}$ and free energy, $E_{\rm free}$, for individual time snapshots with no cycle averaging. Each symbol denotes a run with different emerging region properties, separately for (a) runs where all regions have the same twist and (b) runs with varying amplitudes and signs of twist.}
    \label{fig:scatter-snaps}
\end{figure}

Figure \ref{fig:scatter-snaps} shows that when the time averaging is removed from Figure \ref{fig:scatter}(a), there is more scatter in the relation between unsigned helicity and free energy. For run T0 the relation is still approximately linear for all time snapshots, but when emerging regions are twisted, there is more variation. In this case, the value for run T0 seems to act as an approximate lower bound on the free energy, at a given level of unsigned helicity.

Figure \ref{fig:scatter}(b) shows that the eruptivity seems to depend not only on $\overline{H}$ but also on whether all regions have the same twist (runs TU0.05 and TU0.1) or have varying amplitudes and signs of twist (runs TOb5 and TOb10). Specifically, for a given overall unsigned helicity, there are more eruptions when $\tau$ varies between regions rather than being uniform. This effect is also visible by comparing Figures \ref{fig:cycle-twist}(b) and \ref{fig:cycle-twist}(e). We speculate that the cause could be the distribution of many different magnitudes $|\tau|$ (of both signs), with a sizeable tail of regions having  $|\tau|>0.1$, namely 714 out of 1072 regions in run TOb10, even though the overall average helicity is less than run TU0.1 where all regions have $|\tau|=0.1$. The most twisted of these regions are likely to erupt rapidly, before the magnetic field has a chance to relax or dissipate strong currents.

\subsection{Idealised bipole approximation}

Figures \ref{fig:cycle-twist} and \ref{fig:scatter} also show the run BMR0 where the emerging regions were replaced with idealised bipolar magnetic regions (BMRs). The corresponding values of $\overline{H}$ and $E_{\rm free}$ are approximately $8\%$ and $18\%$ higher than run T0, and this increase is consistent with the increase in surface flux $\Phi_0$ seen in Figure \ref{fig:cycle-phot}(b). In Figure \ref{fig:scatter}(b), we see that despite the increase $\overline{H}$ the eruptivity $\langle\ddot\Phi_1\rangle$ in run BMR0 is comparable to run T0. This suggests that the additional helicity is well distributed and not leading to the more concentrated flux rope structures that would generate more eruptions.

\section{Conclusions}\label{s:conclusions}

In this paper we have presented a picture of how the global coronal magnetic field may have varied over Solar Cycle 24, based on a simplified model that nevertheless retains some of the key physics discarded by the PFSS model.
Our model confirms the finding of \citet{chifu2022} that free energy peaked in late 2014, and we have shown that this is also true of (unsigned) magnetic helicity. It also confirms the result of \citet{yeates2010} that the magneto-frictional model can reproduce the observed pattern of eruptivity varying roughly like the sunspot number. On the other hand, we have shown that this pattern of eruptivity relies on active regions emerging twisted (i.e., with magnetic helicity). This is consistent with previous modelling of active region formation by emerging magnetic flux tubes, which are found to emerge more readily if they are twisted \citep{cheung2014}. Interestingly, \citet{mackay2022} show that one can obtain similar coronal energisation if there is sufficient small-scale helicity injection which accumulates along PILs in the mean field.

We caution that the actual numbers obtained for many of the quantities in our study are still likely to be model dependent. The comparison study by \citet{yeates2018} makes clear that different non-potential models have widely differing $E_{\rm free}/E_{\rm p}$, due to the inclusion of different forms of electric current. Even within the scope of the magneto-frictional model, \citet{mackay2022} have shown how different formulations of the non-ideal term ${\bm N}$ can have a substantial effect on the free energy, with fully ideal simulations (${\bm N}={\bm 0}$) having $E_{\rm free}/E_{\rm p}$ in excess of $60\%$. Compared to this, our simulations are rather conservative in the amount of free energy (the cycle-average ratio $E_{\rm free}/E_{\rm p}$ is about $20\%$ in TU0.05 or TOb5, and about $25\%$ in TU0.1 or TOb10). \citet{mackay2022} also showed that assimilating active regions at their time of maximum flux rather than central meridian crossing leads to substantial increases in $\Phi_0$ and $E$, though not always in $E_{\rm free}/E_{\rm p}$. 

Another problem is numerical resolution. Appendix \ref{s:resolution} shows that halving the resolution of our simulations still gives reasonable estimates of helicity and free energy in 2014, though it underestimates eruptivity. But in general, the NLFFF modelling of active regions shows that extrapolations at higher resolution typically contain more free energy \citep{derosa2015,thalmann2022}.

In future, more stringent calibration against observations will likely help to choose between the different models. Other than the surface magnetic field \citep[which was calibrated by][]{yeates2020}, this was beyond the scope of this study because (a) the comparison is necessarily indirect owing to the lack of coronal magnetic observations, so it is difficult to be quantitative; and (b) optimising the coronal parameters in the model is computationally very expensive. Therefore, whilst our model is based on physical principles, it is still only indicative of likely trends. Possible contraints that have previously been used -- albeit not for a full solar cycle -- include sheared or twisted fields in filament channels \citep{yeates2008, mackay2018, mackay2022}, the shapes of coronal streamers in extreme-ultraviolet \citep{meyer2020, mackay2022, wagner2022}, the locations of coronal holes \citep{yeates2018}, the total open/heliospheric flux \citep{linker2017}, or even the timing of specific eruptions \citep{yardley2021a}. We hope to pursue this in future. 


\begin{acks}
This work was supported by UK STFC grants ST/W00108X/1 and ST/V00235X/1. The SDO data are courtesy of NASA and the SDO/HMI science team. Figures \ref{fig:em-demo} and \ref{fig:corona} used PyVista \citep{sullivan2019}. The author thanks the anonymous reviewer for useful suggestions.
\end{acks}

%



\begin{dataavailability}
The database of emerging regions is available as a pre-generated ascii file on the Harvard Dataverse \citep{yeatesdata2020}. This may be reproduced with either the same or different parameters using our open-source Python code at \url{https://github.com/antyeates1983/sharps-bmrs}. The author will be happy to share data from the magneto-frictional simulations upon reasonable request.
\end{dataavailability}

%

\appendix

\section{Unsuitability of relative helicity} \label{s:hr}

In solar physics, the relative helicity, $H_R$, of \citet{berger1984} is frequently used as a measure of global magnetic topology. Like the unsigned helicity $\overline{H}$ used in this paper, it is unchanged under ideal deformations within the domain that fix the field line footpoints on the boundary. Unlike $\overline{H}$, it has the advantage of being a gauge invariant quantity, but it replaces gauge dependence by dependence on a choice of reference field. It is conventional, however, to choose a potential field as the reference.

\begin{figure}
    \centering
    \includegraphics[width=\textwidth]{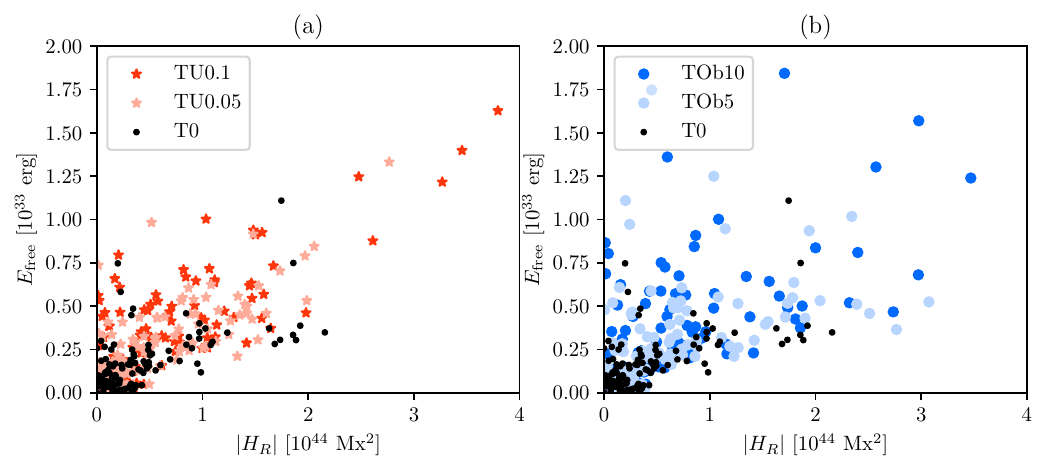}
    \caption{Relation between (unsigned) relative helicity, $H_R$ and free energy, $E_{\rm free}$, for individual time snapshots with no cycle averaging. As in Figure \ref{fig:scatter-snaps}, Each symbol denotes a run with different emerging region properties, separately for (a) runs where all regions have the same twist and (b) runs with varying amplitudes and signs of twist.}
    \label{fig:scatter-hr}
\end{figure}

Unfortunately, $H_R$ is not a useful measure for global simulations of the solar corona, because it is a signed quantity that suffers from cancellation between positive and negative values of the integrand. These can come from regions that are remotely located on the Sun and do not interact in reality. To illustrate this, Figure \ref{fig:scatter-hr} repeats Figure \ref{fig:scatter-snaps}, but using $|H_R|$ on the horizontal axis instead of $\overline{H}$. There is greatly increased scatter compared to Figure \ref{fig:scatter-snaps}, and, in particular, $E_{\rm free}$ does not go to zero as $|H_R|$ goes to zero, unlike the behaviour with $\overline{H}$. This highlights how a low value of $|H_R|$ can result from cancellation of opposite signs and need not indicate a corona near to potential.

Relative helicity is most commonly computed using the \citet{finn1985} formula,
\begin{eqnarray}
    H_R = \int_V({\bf A} + {\bf A}_{\rm ref})\cdot({\bf B} - {\bf B}_{\rm ref})\,\mathrm{d}V,
    \label{eqn:finn}
\end{eqnarray}
where ${\bf B}_{\rm ref} = \nabla\times{\bf A}_{\rm ref}$ is the reference field. Since $H_R$ is gauge invariant for both ${\bf A}$ and ${\bf A}_{\rm ref}$, if we use the canonical gauge ${\bf A}={\bf A}^*$ from Section \ref{s:nonpot} and choose the gauge of ${\bf A}_{\rm ref}$ such that ${\bf A}^*\times{\bf A}_{\rm ref}\cdot{\bf n} = 0$ on $\partial V$, then we can write
\begin{equation}
    H_R = \int_V{\bf A}^*\cdot{\bf B}\,\mathrm{d}V = \frac12\int_{\partial V}\mathcal{A}|B_r|\,\mathrm{d}S,
    \label{eqn:hr}
\end{equation}
where the second equality requires all field lines to intersect the boundary so as to be counted in the integral \citep{berger1988}. Note that if ${\bf B}\equiv{\bf B}_{\rm ref}$ then the integrand in (\ref{eqn:finn}) vanishes everywhere, while in (\ref{eqn:hr}) the integral vanishes only after integrating over all field lines.

\begin{figure}
    \centering
    \includegraphics[width=\textwidth]{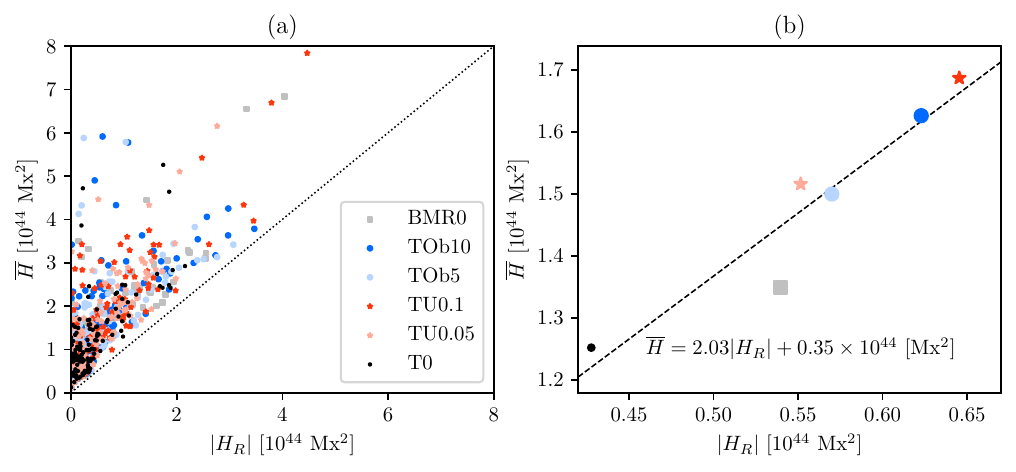}
    \caption{Scatter plots between unsigned helicity, $\overline{H}$, and (unsigned) relative helicity, $H_R$, for (a) individual time snapshots, and (b) cycle-averaged values in each simulation run. The dashed line in (b) shows the indicated linear fit, using all six points.}
    \label{fig:hr-vs-hbar}
\end{figure}

Comparing Equation (\ref{eqn:hr}) with Equation (\ref{eqn:hbar}) shows that $|H_R|\leq \overline{H}$, and this is clearly seen in Figure \ref{fig:hr-vs-hbar}(a), which plots $\overline{H}$ against $|H_R|$ for individual time snapshots. It is very clear from this figure that the two measures are not correlated at individual times, and also the fluctuation of $|H_R|$ makes its change from Minimum to Maximum less meaningful. On the other hand, there is a reasonable correlation between $|H_R|$ and $\overline{H}$ when averaged over all snapshots during the cycle (Figure \ref{fig:hr-vs-hbar}b), albeit with some scatter -- for example, run BMR0 has lower than expected $\overline{H}$ given $|H_R|$, indicating a more uniform sign of $\mathcal{A}$ owing to the simpler surface $B_r$ pattern. This correlation persists in Figure \ref{fig:scatter} if $\overline{H}$ is replaced by $|H_R|$ (not illustrated), albeit weaker. Taking the linear fit for $\overline{H}$ against $|H_R|$ from Figure \ref{fig:hr-vs-hbar}(b) -- with its significant non-zero offset -- and inserting this into the power law from Figure \ref{fig:scatter}(a), suggests an offset power law $|H_R|\,[\mathrm{Mx}^2] = 3.58\times 10^{31}E_{\rm free}^{0.38} - 0.17\times 10^{44}\,[\mathrm{erg}]$.

\section{Parameter dependence}\label{s:params}

\begin{figure}
    \centering
    \includegraphics[width=\textwidth]{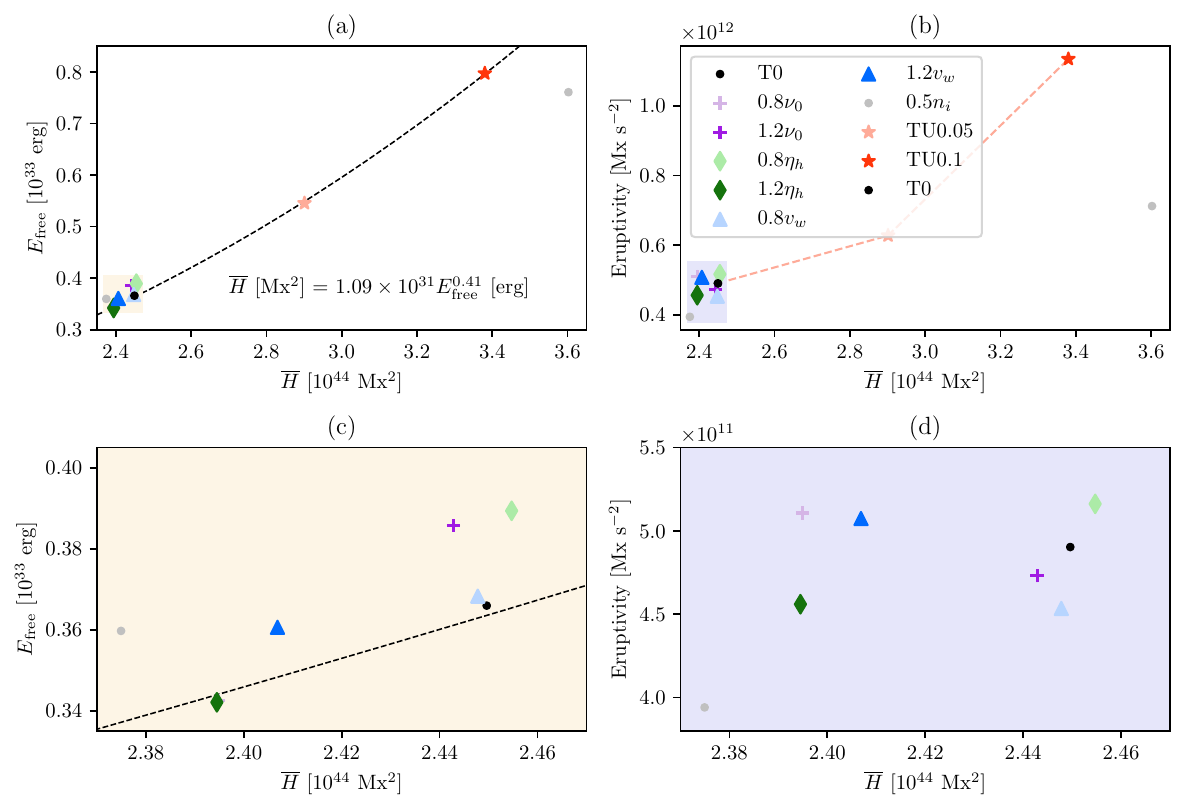}
    \caption{Parameter dependence of the relations between averaged unsigned helicity, $\overline{H}$ and (a) free energy, $E_{\rm free}$; (b) eruptivity, $\langle\ddot\Phi_1\rangle$, computed over 2014 (see text). Panels (c) and (d) show enlargements of the regions around run T0. The dashed line in (a) shows a power-law fit between T0, TU0.05 and TU0.1 only, for context, while in (b) the dashed line guides the eye between these three runs.}
    \label{fig:scatter-params}
\end{figure}

The parameters in the magneto-frictional model are not directly constrained by observations. Therefore, to check the robustness of our results, we have carried out further 12-month runs, varying one coronal parameter at a time compared to run T0. Each of these runs was initialised using the same three-dimensional snapshot of the non-potential ${\bm B}$ from run T0 on 2013 December 29. Figures \ref{fig:scatter-params}(a,b) show an analogous plot to Figure \ref{fig:scatter}, but with the quantities $\overline{H}$, $E_{\rm free}$ and $\langle\ddot\Phi_1\rangle$ averaged only over the 12 months of 2014. The parameters which have been varied are indicated in the legend: we try both reducing and increasing by $20\%$ the friction coefficient $\nu_0$, the hyperdiffusivity $\eta_h$, and the outflow speed $v_w$. In all of these cases, the other parameters keep their default values and the solar surface $B_r$ and emerging region properties remain identical. The figure also shows runs T0, TU0.1 and TU0.5, and notice that the average values are all increased compared to Figure \ref{fig:scatter} because the averages are taken over 2014 only (an active period), rather than the full simulation. Note that the exponent of the power-law fit in Figure \ref{fig:scatter-params}(a) is comparable to that for the whole cycle in Figure \ref{fig:scatter}.

The most significant conclusion is that the $20\%$ parameter variations do not substantially affect the mean quantities in Figure \ref{fig:scatter-params}, leading to standard deviations (relative to run T0) of only $1\%$ in $\overline{H}$, and $5\%$ in $E_{\rm free}$ or $\langle\ddot\Phi_1\rangle$. This is much smaller than the effect of twisting the emerging regions, as seen by comparing runs TU0.05 or TU0.1.

Looking more closely, Figures \ref{fig:scatter-params}(c,d) show enlargements of the region around T0 in Figures \ref{fig:scatter-params}(a,b), respectively. Firstly, note that increasing/decreasing the friction coefficient $\nu_0$ has the effect of increasing/decreasing $E_{\rm free}$, because higher $\nu_0$ means slower relaxation.
On the other hand, increasing/decreasing $\eta_h$ reduces/increases $E_{\rm free}$, because the hyperdiffusion acts to dissipate magnetic energy. Varying $v_w$ has only a weak effect on $E_{\rm free}$, because most of the energy is stored at lower heights which are weakly affected by the outflow. On the other hand, $v_w$ has a relatively stronger effect on the eruptivity in Figure \ref{fig:scatter-params}(d), with higher $v_w$ increasing $\langle\ddot\Phi_1\rangle$ and \textit{vice versa}. Increasing/decreasing $\eta_h$ reduces/increases $\langle\ddot\Phi_1\rangle$, which is consistent with \citet{yeates2009} who found that decreased ohmic diffusion leads to more flux rope eruptions because ropes that form are larger and more twisted. Finally increasing/decreasing $\nu_0$ leads to a small reduction/increase in $\langle\ddot\Phi_1\rangle$ consistent with the change in $E_{\rm free}$.

\subsection{Mesh resolution} \label{s:resolution}

There are two additional points in Figure \ref{fig:scatter-params} labelled $0.5n_i$. These represent completely new simulation runs (from 12 June 2010) at half of the original mesh resolution, namely $(30, 90, 180)$ cells, firstly with $\tau=0$ and secondly with uniform twist $|\tau|=0.1$. In these runs, the emerging regions were re-extracted from HMI/SHARPs, and there are fewer of them because some of the original regions fell below the resolution limit. Nevertheless, Figure \ref{fig:scatter-params}(a) shows that these much less computationally intensive runs were still able to predict the 2014 averages of $E_{\rm free}$ to within $5\%$ for both T0 and TU0.1, and of $\overline{H}$ to within $\pm 7\%$. On the other hand, Figure \ref{fig:scatter-params}(b) shows that the eruptivity $\langle\ddot\Phi_1\rangle$ is underestimated by $20\%$ for run T0 and almost $40\%$ for run TU0.1.

%

\bibliographystyle{spr-mp-sola}
\bibliography{yeates}  

%
%
%

\end{article} 
\end{document}